\documentclass[8.5pt,twoside,twocolumn]{article}
\usepackage[super,sort&compress,comma]{natbib}
\usepackage{mhchem}
\usepackage{times,mathptmx}
\usepackage{sectsty}
\usepackage{balance}

\usepackage{graphicx} 
\usepackage{lastpage}
\usepackage[format=plain,justification=raggedright,singlelinecheck=false,font=small,labelfont=bf,labelsep=space]{caption}
\usepackage{fancyhdr}
\begin{document}

\thispagestyle{plain}
\fancypagestyle{plain}{
\renewcommand{\headrulewidth}{1pt}}
\renewcommand{\thefootnote}{\fnsymbol{footnote}}
\renewcommand\footnoterule{\vspace*{1pt}%
\hrule width 3.4in height 0.4pt \vspace*{5pt}}
\setcounter{secnumdepth}{5}

\makeatletter
\def\subsubsection{\@startsection{subsubsection}{3}{10pt}{-1.25ex plus -1ex minus -.1ex}{0ex plus 0ex}{\normalsize\bf}}
\def\paragraph{\@startsection{paragraph}{4}{10pt}{-1.25ex plus -1ex minus -.1ex}{0ex plus 0ex}{\normalsize\textit}}
\renewcommand\@biblabel[1]{#1}
\renewcommand\@makefntext[1]%
{\noindent\makebox[0pt][r]{\@thefnmark\,}#1}
\makeatother
\renewcommand{\figurename}{\small{Fig.}~}
\sectionfont{\large}
\subsectionfont{\normalsize}

\fancyfoot{}
\fancyhead{}
\renewcommand{\headrulewidth}{1pt}
\renewcommand{\footrulewidth}{1pt}
\setlength{\arrayrulewidth}{1pt}
\setlength{\columnsep}{6.5mm}
\setlength\bibsep{1pt}

\twocolumn[
  \begin{@twocolumnfalse}
\noindent\LARGE{\textbf{Grain refinement and partitioning of impurities in the
 grain \\boundaries of a colloidal polycrystal}}
\vspace{0.6cm}

\noindent\large{\textbf{Neda Ghofraniha,$^{\ast}$\textit{$^{a,b\ddag}$} Elisa Tamborini,\textit{$^{a,b}$} Julian Oberdisse,\textit{$^{a,b}$} Luca Cipelletti,\textit{$^{a,b}$}
and Laurence Ramos\textit{$^{a,b}$}
}}\vspace{0.5cm}

\noindent\textit{\small{\textbf{Received Xth XXXXXXXXXX 20XX, Accepted Xth XXXXXXXXX 20XX\newline
First published on the web Xth XXXXXXXXXX 200X}}}

\noindent \textbf{\small{DOI: 10.1039/b000000x}}
\vspace{0.6cm}

\noindent \normalsize{We study the crystallization of a colloidal
model system in presence of secondary  nanoparticles  acting as
impurities. Using confocal microscopy, we show that the
nanoparticles segregate in the grain boundaries of the colloidal
polycrystal. We demonstrate that the texture of the polycrystal can
be tuned by varying independently the nanoparticle volume fraction
and the crystallization rate, and quantify our findings using
standard models for the nucleation and growth of crystalline materials. Remarkably, we find that the efficiency
of the segregation of the nanoparticles in the grain-boundaries is
determined solely by the typical size of the crystalline grains. }
\vspace{0.5cm}
\end{@twocolumnfalse}
]

\section{Introduction}


\footnotetext{\textit{$^{a}$~Universit\'{e} Montpellier 2, Laboratoire Charles Coulomb UMR 5221, F-34095, Montpellier,
France. E-mail: neda.ghofraniha@roma1.infn.it}}
\footnotetext{\textit{$^{b}$~CNRS, Laboratoire Charles Coulomb UMR 5221, F-34095, Montpellier, France.}}


\footnotetext{\ddag~Present address: IPCF-CNR, Dipartimento di Fisica, Universit$\grave{\mathrm a}$ La Sapienza, Rome-Italy.}

Impurities affect drastically crystal nucleation, growth and texture and understanding
the role they play during the solidification processes is of great interest in
material science and engineering~\cite{Pan00}, pharmaceutical industry~\cite{Yul04},
mineralogy~\cite{Dav00}, protein crystallization~\cite{Che03, Mat06}  and life science~\cite{Dev06}.

Moreover, the control of the texture by addition of solutes is important for tailoring the
mechanical properties of crystals~\cite{Bub06,Ley10,Sch98}.

Relevant progress in understanding crystallization  of mono and
bidisperse systems has been achieved using colloids, that are often
regarded as an analog to atoms on larger length and time
scales~\cite{Her10}, enabling real time studies by direct
visualization of processes of importance and interest in condensed
matter physics~\cite{gasser01,Sch04}. Crystallization of nano- and
micro-spheres have been widely investigated
numerically~\cite{auer05} and
experimentally~\cite{Pal99,McC96,Yue96}, including in bi-component
systems. Binary mixtures of colloids where the volume fraction of
the two species is not too unbalanced (up to a factor of 7, but most
typically by a factor of 2) exhibit a great variety of crystalline
superlattice structures similar to atomic
systems~\cite{Bartlett92,Schofield05,Hac80,Leu05}.
By contrast, the addition of a very small amounts of a dopant in a
solidifying matrix does not change the crystalline unit cell, but
can modify the nucleation~\cite{Cac04,Sea06,Engelbrecht10}, and induce
crystal growth frustration~\cite{deV05Sci,Yoshi2011}, local defect
formation~\cite{deV09SM} and local fractionation~\cite{Mar03},
depending on the volume fraction and size of the dopants.

In this article we present a novel strategy for the control of the
texture of colloidal polycrystals, based on the addition of small
amounts of nanoparticles (NPs) as dopants to a solidifying matrix.
We  achieve the segregation of NPs in a network of thin sheets that
is the colloidal counterpart of the texture of grain boundaries
formed during the solidification of crystalline metallic
alloys~\cite{Lee00} and other molecular materials doped with
impurities~\cite{Los95}. By varying the NP volume fraction and the
rate at which the background matrix is solidified, both the typical
mesh size of the dopant network and the efficiency of the
partitioning of the NPs between the matrix and the grain boundaries
are tuned. In the framework of the classical nucleation theory, we
show quantitatively for the first time how both the nucleation and
the growth of colloidal crystallites are influenced by the NP volume
fraction and the crystallization rate, determining the final grain
size.

\section{Materials and methods}
\label{sec:methods}

The colloidal polycrystal is composed of an aqueous solution of
Pluronic F108, a commercial PEO-PPO-PEO triblock co-polymer (Serva
Electrophoresis GmbH), where PEO and PPO denote polyethylene oxide
and polypropylene oxide, respectively. The co-polymer (concentration
of 34\% w/w) is fully dissolved at  ${ T \approx
0~^{\circ}\mathrm{C}}$. Upon increasing $T$, it self-assembles into
spherical micelles of diameter $d\sim 22$ nm, due to the increased
hydrophobicity of the PPO block~\cite{Pas95}. The volume fraction,
$\phi$, of the micelles increases with $T$, until crystallization
occurs due to micelle crowding. Differential scanning calorimetry
and rheology show that crystallization occurs between
$15~^{\circ}\mathrm{C}$ and $16~^{\circ}\mathrm{C}$, depending on
the heating rate~\cite{AmeurInPreparation}. We emphasize that,
unlike conventional colloidal systems where controlling precisely
and changing \emph{in situ} the volume fraction is a difficult task,
our system allows crystallization to be induced at the desired rate
simply by varying $T$, in analogy with molecular materials.

In order to analyze in a quantitative way the crystallization
process, as it will be discussed in Sec.~\ref{sec:resdisc}, it is
convenient to map our micellar system onto a hard sphere suspension,
by determining the relation between $T$ and $\phi$. The soundness of
this approach is supported by previous work~\cite{Mortensen92}
showing that the static structure factor of a similar micellar
system could be mapped onto that of hard spheres using an affine
relation between $T$ and $\phi$. In our case, we determine $\phi(T)$
by comparing the temperature dependence of the sample viscosity in
the micellar fluid phase to the volume fraction dependence of the
viscosity of hard-spheres suspensions~\cite{ChaikinPRE}. The
mapping thus determined is $\phi = \alpha \times (T-T_0)$ with
$T_0=4.2~{^\circ}\mathrm{C}$ and $\alpha \simeq
0.045~{^\circ}\mathrm{C}^{-1}$~\cite{AmeurInPreparation}. With
this mapping, rheology measurements in the regime where the sample
is solidified very slowly so as to avoid any significant
undercooling show that the onset of crystallization occurs at $\phi
 = \phi_{\rm cryst} \approx 0.5136$, a value compatible with that
expected for slightly polydisperse hard spheres.

We dope the micellar crystal with small amounts (at most $1\%$ v/v)
of green-yellow fluorescent carboxylated polystyrene nanoparticles
purchased from Invitrogen, with diameter $\sigma = 20, 36$ and $100$
nm.  Neutron scattering measurements show that the microscopic
crystalline structure (face-centered cubic, with lattice parameter
$30$ nm) is preserved upon addition of up to 2\% v/v of NPs
\cite{Tamborini12}, twice than the highest concentration used here.
For confocal microscopy imaging, samples are introduced in chambers
of thickness $250 \mu$m and solidified by raising the temperature
from $3^\circ$C to $23^\circ$C at a controlled rate, $\dot{T}$, and
visualized at least $20  \mu$m from the walls.

\section{Results and Discussions}
\label{sec:resdisc}

\begin{figure}[h]
\centering
  \includegraphics[height=8cm]{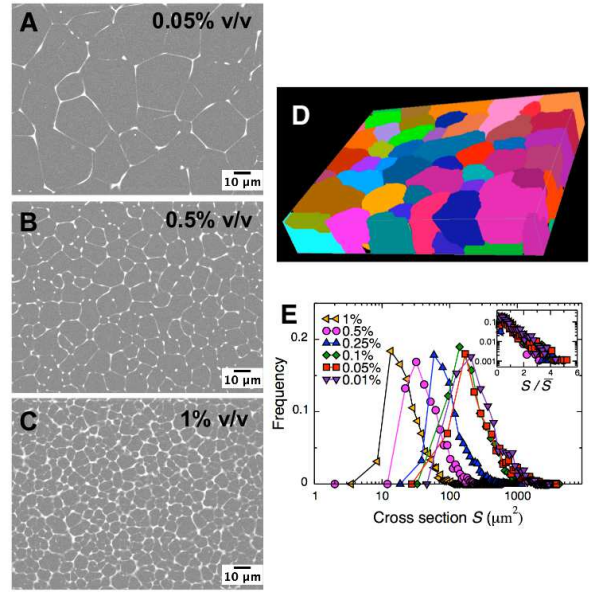}
  \caption{ Color online. (A-C)  Confocal
microscopy images of three samples prepared at different impurities
(NP) volume fraction $c_ 0$. The clear lines are sections in the observation
plane of the grain boundaries, where the NPs accumulate. The
temperature ramp is $\dot{T} = 7 \times 10 ^{\mathrm
-3}~{^\circ}\mathrm{C~min}^{\mathrm -1}$. (D) 3-D reconstruction of
a sample with $c_0=0.5\%$. The size of the reconstructed region is
35$\times$30$\times$5 ~$\mu$m$^ 3$. (E) Frequency distribution of
the cross section of the grains, for samples  with various $c_0$
prepared at a constant $\dot{T}$ = 7$\times$10$^{\mathrm
-3}~{^\circ}\mathrm{C~min}^{\mathrm -1}$. Inset: same data as in the
main plot but scaled by the average cross section $\overline{S}$.
Impurities are fluorescently labeled polystyrene NPs with diameter
$36$ nm.}
  \label{fig:1}
\end{figure}

We show in
Figs.~\ref{fig:1}A-C representative images of
colloidal samples doped with NPs with diameter $\sigma
= 36~\mathrm{nm}$. All samples are solidified at the same
rate, $\dot{T} = 7 \times
10^{-3}~{^{\circ}}\mathrm{C}~\mathrm{min}^{-1}$, and the average NP
volume fraction $c_0$ is varied
between $0.05\%$ and $1\%$. Images exhibit fluorescent (clear) connected thin
lines that separate featureless (dark) zones. Confocal scans
through the material show that the clear lines are in fact 2-D
sections of a network of grain boundaries enriched in NPs that
delimitate grains with different crystalline orientations.
A 3-D reconstruction of sample with $c_0$=$0.5\%$ is reported in Fig.~\ref{fig:1}D,
where different false colors indicate different grains.
Figures~\ref{fig:1}A-C clearly show that the grain size
 decreases by increasing the amount of NPs.
To quantify the influence of NPs, we calculate the cross section
area of the grains, $S$. The frequency distribution of $S$ for
various $c_0$ is reported in Fig.~\ref{fig:1}E [about $800$ cross
sections, taken in distinct positions on both sides of the chamber
are measured for each experimental condition]. The peak of the
distribution moves to smaller values with increasing $c_0$,
demonstrating grain refinement at larger impurity content.

\begin{figure}[h]
\centering
  \includegraphics[height=6cm]{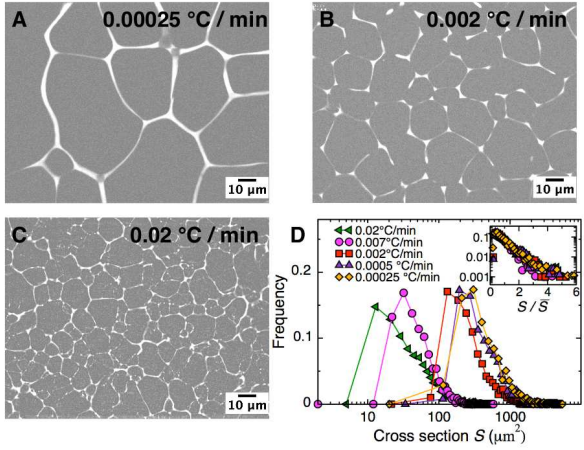}
  \caption{Color online. (A-C)  Confocal
microscopy images of three samples prepared at various $\dot{T}$.
The NP volume fraction is fixed at $c_0=0.5\%$. (D) Frequency
distribution of the cross section of the grains, for samples with
$c_0=0.5\%$ and prepared with various $\dot{T}$. Inset: same data as
in the main plot but scaled by the average cross section
$\overline{S}$. Impurities are fluorescently labeled polystyrene NPs
with diameter $36$ nm.}
  \label{fig:2}
\end{figure}

In analogy to atomic crystals and other materials \cite{Sharma2011},
we find that the rate at which temperature is changed during
solidification has also a crucial influence on the final grain size,
slower ramps leading to a coarser texture. This is demonstrated for
samples with a fixed NP volume fraction ($c_0 = 0.5\%$) and prepared
with various temperature ramps, as in the images in
Figs.~\ref{fig:2}A-C and in figure~\ref{fig:2}D
 where the size distributions of
the cross section area of the grains, spanning two decades, are
reported. In all cases, the distributions are relatively narrow,
their standard deviation normalized by the average being $0.7 \pm
0.2$. Remarkably, we find that the shape of the distributions is
almost independent of the NPs volume fraction and of the temperature
ramp, as demonstrated by the inset of Fig.~\ref{fig:1}E (resp.
Fig.~\ref{fig:2}D), where curves for various  $c_0$ (resp.
$\dot{T}$) are collapsed by normalizing the $S$ with respect to its
mean value, $\overline{S}$. In the insets, the right tails of the
distributions are approximately straight lines in a semi-logarithmic
plot, indicating exponential tails.

\begin{figure}[h]
\centering
  \includegraphics[height=3.5cm]{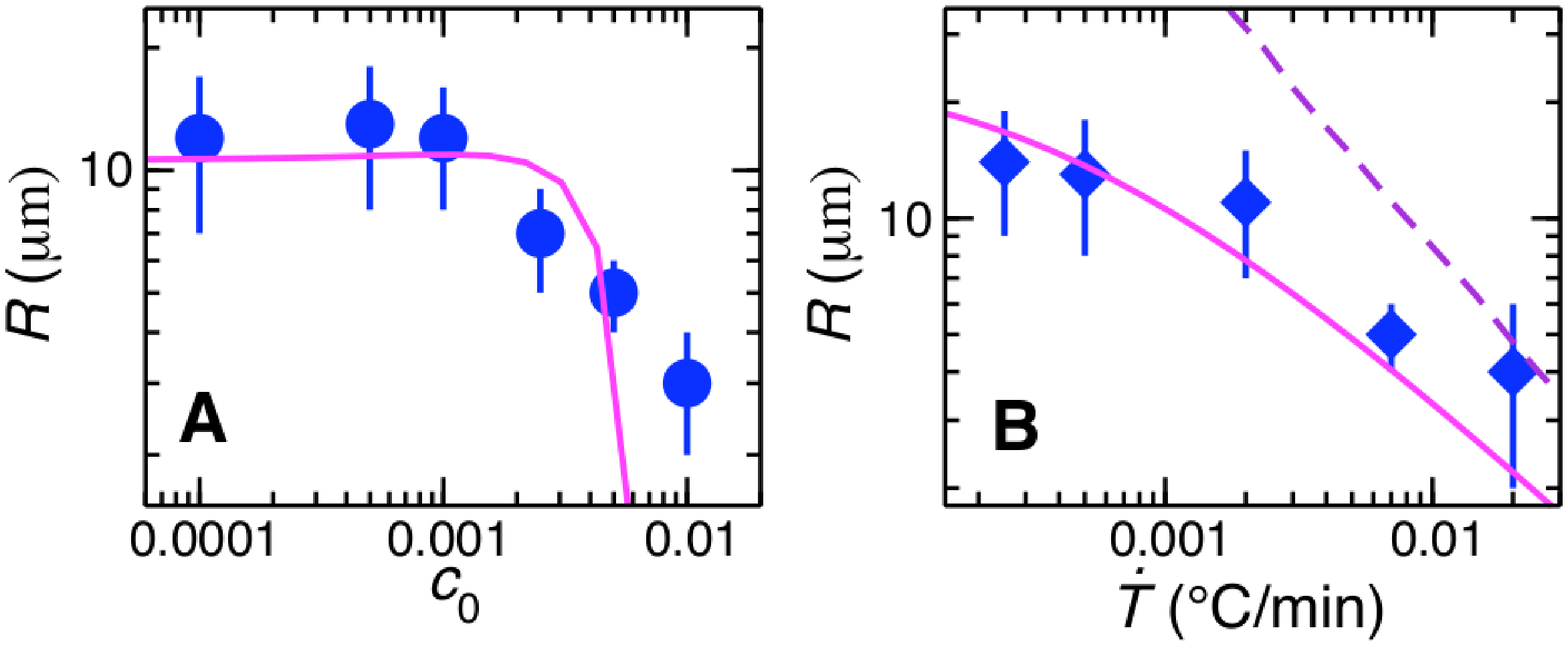}
  \caption{
 (A) Mean radius of the grains $vs.$
 volume fraction of impurities, at a fixed temperature ramp $\dot{T}$=
7$ \times {\mathrm10}^{\mathrm -3}~{^{\circ}}\mathrm{C~min}^{\mathrm -1}$ and
 (B) $vs.$ $\dot{T}$, at a fixed impurity volume fraction $c_0$=0.5\%. Symbols are
 experimental data and lines are numerical calculations, as explained in the text.
 The dashed line is the theoretical $\dot{T}$ dependence in the absence of impurities.
 In A and B the bars are the standard deviation of the size distributions.}
\label{fig:cross_section}
\end{figure}

Figures~\ref{fig:cross_section}A and B display the evolution of the
mean radius of the grain, $R= \sqrt{\overline S/\pi}$, where
$\overline S$ is the mean cross-section area, with impurity amount
$c_{0}$ and temperature ramp  $\dot{T}$, respectively. The grain
size is of the order of a few micrometers$^{\dagger}$, with a
clear influence of both NP content and crystallization on the size.

\footnotetext{$^{\dagger}$ Note that strictly speaking $R$ is smaller than the actual average
grain radius, since $S$ is the grain cross section obtained by a
random slicing, not the maximal cross section. However, we expect
the difference to be relatively small: for an isolated sphere, e.g.,
the average radius of a random cross section differs from the actual
sphere radius by less than $20\%$.}

To quantitatively analyze our experimental findings, we use
classical models for the nucleation and growth of crystalline
grains. We first recall how an average grain size can be computed,
using the Johnson-Mehl-Avrami-Kolmogorov (JMAK) theory
\cite{Kolmogorov1937,Johnson1939,Avrami1939,Avrami1940,Avrami1941}
which assumes that nucleation occurs randomly and homogeneously, and
that the growth velocity of the crystallites, $v_g$, does not depend
on the extent of the crystallization process and is isotropic.
Following JMAK, the extended volume fraction of the sample that has
crystallized at time $t$, $X_e(t)$, reads:
\begin{equation}
X_{e}(t) =  \frac{4 \pi}{3} \int_{0}^{t}  I(\tau)  \left [ \int_{\tau}^{t}v_g(\tau')\mathrm{d}\tau'\right]^3  \mathrm{d}\tau
\label{eq:X_ext}
\end{equation}
where $I(\tau)$ is the nucleation rate per unit volume at time
$\tau$. The extended volume fraction does not take into account that
as crystallization proceeds less fluid phase is available for
nucleating new crystallites. Moreover, it doesn't account for the
fact that crystallites stop growing when they impinge into each
other. As shown by Kolmogorov and Avrami, these effects can be
accounted for by calculating the actual volume fraction that has
crystallized, $X_a(t)$, according to :
\begin{equation}
X_a(t) =  1-\exp[-X_e(t)] \label{eq:JMAK}
\end{equation}

The expected grain radius $R$  can be calculated from the grain
number density at the end of the crystallization
process~\cite{Farjas2008}, yielding
\begin{equation}
R =  \left ( \frac{4 \pi}{3} \int_{0}^{\infty}  I_a(\tau)
\mathrm{d}\tau \right ) ^{-1/3} \label{eq:R}
\end{equation}
where $I_a$ is the actual nucleation rate:
\begin{equation}
I_a(\tau) =  \left [ 1- X_a(\tau)\right ] I(\tau) \label{eq:I_a}
\end{equation}

Note that the nucleation rate, $I$, and growth rate, $v_g$, are in
general time-dependent quantities. This is indeed the case in our
experiments, where crystallization occurs while the sample is
submitted to a continuous increase of temperature, leading to a
continuous increase of the micelle volume fraction. In the
following, we aim at calculating $R$ using
Eqs.~(\ref{eq:X_ext}-\ref{eq:I_a}) in order to compare it to the
experimental results of Fig.~\ref{fig:cross_section}A-B. To this
end, explicit expressions are required for $I$ and $v_g$. We start
by discussing the growth velocity and assume that $v_g$ is given by
the Wilson-Frenkel law~\cite{auer05}:
\begin{equation}
v_g=\frac{D_s}{\lambda}[1-\exp(-|\Delta\mu|/k_BT)] \,
\label{eq:Vg}
\end{equation}
where $k_\mathrm{B}$ is Boltzmann's constant, $\Delta\mu$ is the
difference in chemical potential between the solid and the liquid
phases, $D_s$ is the self diffusion coefficient of the micelles in
the liquid phase, and $\lambda$ is a typical distance over which
diffusion occurs when a micelle in the fluid phase is incorporated
in the growing crystallite.

To evaluate $\Delta \mu$, we take advantage of numerical and
experimental results on the crystallization of hard spheres (HS)
suspensions~\cite{auer05,Pal99,Yue96,Har97}, assuming that micelles
in first approximation behave as HS and using the $\phi(T)$ mapping
discussed in Sec.~\ref{sec:methods}. We evaluate $\Delta \mu$ from
numerical simulations of HS colloids~\cite{auer05}, where data can
be well fitted by $\Delta \mu = A_{\mu} k_B T  (\phi_{\rm cryst} -
\phi)$ with $\phi_{cryst} \approx 0.5$ and $A_{\mu}\approx 15$. In
the following, we take $\phi_{\rm cryst}= 0.5136$ as discussed in
Sec.~\ref{sec:methods}, while $A_{\mu}$ is treated as an adjustable
parameter. The volume fraction dependence of $D_s$ for the micelles
is estimated from experimental data on the structural relaxation
time associated with self-diffusion in suspensions of HS
colloids~\cite{Brambilla09}, assuming that the diffusion coefficient
is inversely proportional to the structural relaxation time. We use
the empirical formula $D_s=D_0 \times \left [
(1-2.5\phi+1.36\phi^2)(1-\frac{\phi}{\phi_c})+144.57 \phi^2
(1-\frac{\phi}{\phi_c})^{2.5} \right ]$, which reproduces very well
the experimental data in the range $0 \le \phi \le 0.59$, and use
$D_0 = \frac{K_B T}{3 \pi \eta_0 d} = 6 \times 10^{-14} \rm{m}^2
/\rm{s}$ as the diffusion coefficient of a micelle in the dilute
regime, where $\eta_0 =0.32~\mathrm{Pa}~\mathrm{s}$ is the solvent
viscosity. The quantity $\lambda$ is expected to be of the order of
the micelle diameter, $d$, although significative differences
between simulations and experiments have been
reported~\cite{auer05}. We set $\lambda = A_{\lambda} d$ with
$A_{\lambda}$ an adjustable parameter.

We now turn to the nucleation rate $I$. Classical nucleation theory
(CNT)~\cite{ABR-book} predicts
\begin{equation}
I=\Gamma \exp [-\Delta G^\star/k_BT] \,
\label{eq:arrhenius}
\end{equation}
where $\Delta G^\star$ is the nucleation barrier in the Gibbs free
energy to be overcome in order to form a stable nucleus, and the
kinetic prefactor  $\Gamma = (\frac{1}{6 \pi k_BT
n_c}|\Delta\mu|)^{0.5} \rho_L \frac{24 D_s}{\lambda ^2} n_c^{2/3}$.
Here $n_c$ is the number of micelles in the critical cluster, and
$\rho_L=6\phi/\pi d^3$ is the number density of the micelles in the
liquid phase. According to CNT, $\Delta G^\star$ is the maximum in
the Gibbs free energy, which is the sum of a surface free energy
term, $\Delta G_S=  4 \pi r^2 \gamma_0$, and a volume term, $\Delta
G_V= - \frac{4\pi}{3} r^3 \rho_S |\Delta \mu|$. Here $r$ is the
radius of a nucleus, $\gamma_0$ is the surface free energy density
of the fluid-crystal interface and $\rho_S$ is the number density of
micelles in the solid phase ($\rho_S=6\phi_c/ \pi d^3$ with $\phi_c
\cong 0.5$). One readily finds
\begin{equation}
\Delta G^\star = \frac{16 \pi}{3}\frac{\gamma_0^3}{\rho_S|\Delta
\mu}| \,
\label{eq:DeltaGCNT}
\end{equation}
For the surface free energy density, we use $\gamma_0 = k_BT
A_{\gamma}/d^2$ with $A_{\gamma}$ an adjustable parameter expected
to be of order $1$~\cite{gasser01,Har97,Hernandez-Guzman09}.
Numerical simulations of HS~\cite{auer05} show that $n_c$ the number
of colloids in the critical nucleus is of order 100 and depends only
weakly on the volume fraction: accordingly, we take $n_c = 100$
independent of $\phi$.

\begin{figure}[h]
\centering
  \includegraphics[height=9cm]{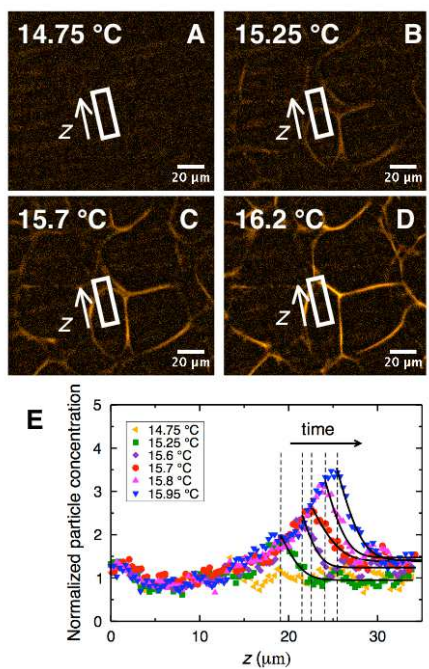}
  \caption{Color online.
(A-D) Confocal microscopy images showing the
progressive accumulation of impurities at the fluid-solid interface
during solidification. Images are taken at different temperatures,
as specified by the labels, during solidification at a rate
{\slshape \.{T}}=0.03~${^\circ}\mathrm{C~min}^{\mathrm -1}$. The impurities are 36 nm
fluorescent NPs, at a concentration of $c_0$=0.05\%. The white
rectangle shows the region of the growing crystallites used for the
analysis of the temporal evolution of the concentration profile
shown below.(E) Time dependence of  the normalized NPs volume fraction
obtained from fluorescence intensity profiles calculated across a
solid-fluid interface.
The doted vertical lines indicate the position of the solid-liquid interface as determined by the peak of the fluorescence profiles.
}
\label{fig:movie}
\end{figure}

In order to model the effect of the addition of NP, we need to
modify the CNT expressions. We assume that the most relevant effect
is a modification of the surface free energy term due to the
accumulation of NP at the interface between the crystal and the
liquid phase (see fig.~\ref{fig:movie}), since slight variations of
$\gamma_0$ are known to have a massive impact on nucleation
rates~\cite{ABR-book}. We thus neglect any dependence of $|\Delta
\mu|$ on NP content, implying that both $v_g$ and the volume term
$\Delta G_V$ are unchanged upon addition of NP, and propose that the
presence of the NPs at the interface leads to a reduction of the
surface free energy density. We assume that this reduction is
proportional to the amount of NPs at the interface, which in turn is
proportional to $n_p$, the amount of NPs initially available in the
volume occupied by a crystallite. Hence, for a nucleus of radius
$r$, the reduction of surface energy is proportional to $n_p=8r^3
c_0 / \sigma^3$, and to the cross-section of the NPs, $\pi \sigma^2
/4$. The surface term then becomes $\Delta G_S = 4 \pi r^2 \gamma_0
- \frac{2\pi \gamma_0 F_{P}c_0}{\sigma} r^3 $, with $F_p$ a
proportionality constant. Note that the NP contribution, although
physically due to a change of the surface term, formally can be
incorporated in the volume term of $\Delta G$, because it scales
with $r^3$. In the presence of NPs, the free energy barrier is then
obtained by a suitable modification of the denominator of
Eq.~(\ref{eq:DeltaGCNT}):
\begin{equation}
\Delta G^\star = \frac{16 \pi}{3} \frac{\gamma_0^3}{[\rho_S \Delta \mu + \frac{3}{2} \gamma_0 F_{P} c_0 / \sigma]^2} \,
\label{eq:DeltaG}
\end{equation}

Using $A_{\mu}$, $A_{\lambda}$, $A_{\gamma}$ and $F_P$ as adjustable
parameters, we fit the experimental $R$ using the modified CNT
approach outlined above. Operationally, we take $t=0$ as the time at
which $\phi = \phi_{\rm cryst}$ and, for a given $\dot{T}$ and
$c_0$, we calculate numerically $X_e(t)$ using Eq.~(\ref{eq:X_ext})
and $v_g(t)$ and $I(t)$ as given by Eqs.
(\ref{eq:Vg},\ref{eq:arrhenius},\ref{eq:DeltaG}), where the latter
quantities depend on time through $D_s$ and $\Delta \mu$. Equations
(\ref{eq:JMAK}-\ref{eq:I_a}) are then used to obtain $R$. We show in
Fig.~\ref{fig:cross_section}A-B the best fits as solid lines. While
not perfect, the fits exhibit a reasonable quantitative agreement
with the experimental data, especially given the order-of-magnitude
variations of quantities related to colloidal crystallization often
observed in the literature~\cite{auer05}. The fitting parameters
are: $A_\lambda = 0.14 \pm 0.05$, $A_\mu =
 17 \pm 3$, $A_\gamma = 0.747\pm 0.20$ and $F_P = 126 \pm 24$. The
three fitting parameters that are independent of NP content and for
which a comparison with previous works is possible are in good
agreement with numerical results~\cite{auer05}, for which $A_\lambda
= 0.2 - 0.5$, $A_\mu \approx 15 $, $A_\gamma = 0.64$. Furthermore,
this simple model is able to capture the main experimental trends,
in particular the roll-off of $R(\dot{T})$ at slow ramps, which is
due to the presence of NPs (compare the continuous and dotted lines
in Fig.~\ref{fig:cross_section}B) . The discrepancy at high $c_0$
observed in in Fig.~\ref{fig:cross_section}A is presumably due to
an over-evaluation of the effect of NPs on the surface tension, due
to the fact that, as we shall show it in the following, as $c_0$
increases the fraction of NPs that are effectively expelled
from a growing crystallite decreases. Although a
qualitative relation between $\Delta G^\star$ and the typical
crystallite size has been reported for metals~\cite{Shi95}, our work
is the first quantitative analysis for colloids of the grain size
using nucleation and growth theories~\cite{auer05}.

\begin{figure}[h]
\centering
  \includegraphics[height=6cm]{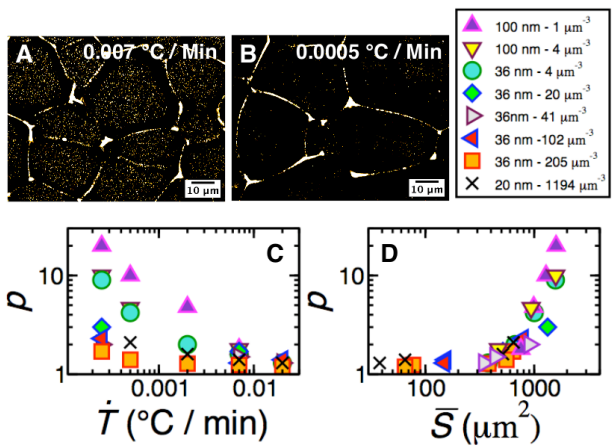}
  \caption{Color online. (A-B) Confocal images of samples prepared
with ramp rates 7$\times$10$^{\mathrm -3}~{^\circ}\mathrm{C~min}^{\mathrm -1}$ and
5$\times{\mathrm 10}^{\mathrm -4}~{^\circ}\mathrm{C~min}^{\mathrm -1}$, respectively.
(C) Ratio $p$
of the volume fraction of NPs in the grain boundaries and inside the
grains. The symbols are labeled by the
particle diameter and their number density, $\rho$. (D)
Same data as in (C)
collapsed onto a master curve
by plotting  $p$ as a function of the average grain cross section.
}  \label{fig:segregation}
\end{figure}

So far, we have shown that both the volume fraction of impurities and the crystallization
rate determine the polycrystalline texture,
thereby allowing for a fine control of the average grain size over almost one decade.
We now turn to the investigation of the degree of confinement of NPs once the solidification process is completed.
To better appreciate the spatial distribution of the NPs, we show
in Figs.~\ref{fig:segregation}A-B
confocal images of samples
doped with particles large enough ($\sigma =  100~\mathrm{nm}$) to be individually
resolved by confocal microscopy, at least for moderate volume fractions.
The dopant volume fraction is kept fixed at $c_0 = 0.05\%$, 
while the effect of $\dot{T}$ is investigated.

As shown in Fig.~\ref{fig:segregation}A, for polycrystals prepared using a relatively fast temperature ramp,
an excess of NPs is observed
in the grain boundaries, consistently with our previous observations with smaller dopants.
However, large numbers of individual NPs are clearly visible also in the interior of the grains.
By contrast, NPs are almost completely expelled from the grains in samples prepared
using a slower ramp (Fig.~\ref{fig:segregation}B).
We quantify the efficiency of the partitioning by evaluating the
ratio, $p$, between the NP volume fraction in the grain boundaries and
that in the interior of the grains, as obtained from fluorescence
intensity measurements. Figure~\ref{fig:segregation}C shows that
confinement in the grain boundaries becomes increasingly efficient
as the temperature ramp slows down, similarly to observations in atomic
systems~\cite{eckler92}, and
that, quite generally, $p$ decreases when the average number density
of impurities, $\rho$, increases. Thus, an
increase of either $\dot{T}$ or $\rho$ yields both a finer texture
and a smaller degree of partitioning, suggesting that the degree of
partitioning should correlate with the grain size.
We note that the grain size, directly related to the number of successful nuclei, does not depend on the size of NPs but only on their quantity, signaling the absence
 of heterogeneous nucleation on the surface of  NPs, as expected since the size of the NPs is comparable to the size of the micelles~\cite{Cac04}.\\
Remarkably, we find
that all data for ramp rates spanning two decades, impurity number
density spanning three decades, and for NP diameters ranging between
20 nm and 100 nm indeed collapse onto a master curve when plotting
$p$ as a function of the average grain cross section
(Fig.~\ref{fig:segregation}D).
Interestingly, $p$ is close to unity
and rather insensitive to $\overline{S}$ until $\overline{S} \approx
800~\mu\mathrm{m}^2$, above which the degree of partitioning
increases steeply.

To understand the origin of the correlation between grain size and
degree of particle segregation, we analyze the growth of the
crystallites during solidification, a quantity not accessible in
metallic systems. This is illustrated in Fig.~\ref{fig:movie} and
Supplementary movie S1.

At low temperature, below the onset of
crystallization, the images are homogeneously grey down to the
smallest length scale that is accessible (1 pixel $=
0.11~\mu\mathrm{m}$), revealing that the impurities are
homogeneously distributed in the sample, which is still fully fluid.
The nucleation and early stage growth can not be resolved in our
confocal microscopy experiments, presumably because the partition of
the fluorescent impurities is too weak (Fig.~\ref{fig:movie}A). As
solidification proceeds, however, the interface of the growing
crystallites becomes increasingly visible, because the fluorescent
NPs are expelled by the solid phase and accumulated in the fluid
boundary layer (Fig.~\ref{fig:movie}B-C). Eventually, the different
grains that are growing touch each other and become facetted,
resulting in the formation of a network of grain boundaries where
the NPs are confined at a density larger than in the bulk
(Fig.~\ref{fig:movie}D). In Fig.~\ref{fig:movie}E we report the
normalized impurity volume fraction profiles, obtained from the fluorescent
intensity, along the growth direction of one grain from its center:
the height of the peak of NP volume fraction increases during the
propagation of the solid-fluid interface, indicating that the amount
of impurities rejected from the crystallite and accumulated at the
interface increases steadily, until the grain stops growing when
impinging on an adjacent growing crystallite. Thus, the larger a
grain can grow, the higher the volume fraction of the impurities
eventually trapped in the grain boundary, explaining the correlation
between the grain size and the degree of partitioning shown in
Fig.~\ref{fig:segregation}D.


As a concluding remark, we note that we have successfully reproduced
our experiments with a variety of NP of different kinds, thereby
demonstrating that the system described here provides a powerful,
robust and convenient way to study the crystallization of
bi-disperse colloidal  systems and to segregate NPs in thin sheets
within a 3-D soft matrix. The colloidal composite material
investigated here, thanks to its softness and optical transparency,
offers a unique opportunity to investigate in real time and by
direct visualization the formation of a polycrystalline texture and
its response to an external perturbation. In particular, we are
currently exploring the dynamics of the grain boundaries under a
mechanical load.

\section*{Acknowledgments}

We thank  P. Olmsted for suggesting the effect of NP on the
nucleation barrier, R. Piazza, M. Cloitre, J.-L. Barrat, and D.
Frenkel for insightful discussions and J. Lambert for the
reconstruction of 3-D images This work has been supported by ANR
under Contract No. ANR-09-BLAN-0198 (COMET)

\footnotesize{
\bibliography{biblio-grain}
\bibliographystyle{rsc}
}

\end{document}